\documentclass{article}
\usepackage{spconf}
\usepackage[T1]{fontenc}
\usepackage[utf8]{inputenc}

\usepackage{amsmath,cite,url}
\usepackage{graphicx}
\usepackage{color}

\usepackage{stfloats} 
\usepackage{caption}
\usepackage{tikz}
\usepackage[inline]{enumitem}
\usepackage{breakcites}
\usepackage{microtype}
\usepackage{xcolor}
\usepackage{array}
\usepackage{float}
\usepackage{adjustbox}
\usepackage{csquotes}
\usepackage{makecell}
\usepackage{pdfpages}
\usepackage{xspace}
\usepackage{subfig}
\definecolor{rank1}{RGB}{112,255,112}
\definecolor{rank2}{RGB}{133,133,133}
\definecolor{rank3}{RGB}{ 69, 69, 69}
\definecolor{rank4}{RGB}{  0,  0,  0}

\usetikzlibrary{positioning,shapes.geometric}

\newcommand{\SIMSESpec}{\texttt{SIMSE\_\allowbreak Spec}\xspace}
\newcommand{\LoneSpec}{\texttt{L1\_\allowbreak Spec}\xspace}
\newcommand{\JTFS}{\texttt{JTFS}\xspace}
\newcommand{\DTWEnv}{\texttt{DTW\_\allowbreak Envelope}\xspace}

\newcommand{\BPNoise}{\textbf{BP-Noise}\xspace}
\newcommand{\BPSaw}{\textbf{BP-Saw}\xspace}
\newcommand{\FMMod}{\textbf{SineSaw-AM}\xspace}
\newcommand{\FMModvtwo}{\textbf{SineSine-AM}\xspace}

\title{Evaluating Loss Functions in Differentiable Out-of-Domain Sound-Matching with Partial Parameter Distance}

\name{Amir Salimi, Daniel Penner, Kalvin Eng, Abram Hindle, Osmar R. Za{\"i}ane}
\address{University of Alberta}

\usepackage[bookmarks=false,hidelinks]{hyperref}

\begin{document}

\maketitle

\begin{abstract}

In \textit{out-of-domain} (OOD) sound-matching, a synthesizer is optimized to mimic a sound it did not generate. OOD evaluation of loss functions is underexplored in part because the standard ``parameter loss'' metric requires a shared parameter space between target and imitator, which OOD settings lack. We introduce Partial Parameter Distance (PPD), which applies parameter loss only to the critical parameters that mismatched synthesizers share (e.g., filter cutoffs), enabling automatically evaluated OOD experiments; we verify its results with blinded listening tests. Across seven scenarios involving band-pass filtering, amplitude modulation, and pitch-bending, we evaluate four differentiable loss functions (\SIMSESpec, \LoneSpec, \JTFS, \DTWEnv). Loss-function effectiveness remains tightly coupled to the method of synthesis: \SIMSESpec excels at filter-cutoff recovery, \DTWEnv at amplitude-modulation recovery, and \JTFS at smooth pitch trajectories. Parameter-based evaluation agrees with listening tests on the top-ranked loss function in five of seven scenarios, demonstrating its utility as a diagnostic tool.

\end{abstract}

\section{Introduction}

In sound design, artists iteratively manipulate the parameters of their digital synthesizers to produce a desired sound~\cite{krekovic2019insights}. This manual process can be a rewarding creative outlet, but often requires expertise and patience. \textit{Sound-matching} refers to the automation of this workflow: given a target sound and a synthesizer, the objective is to identify parameters that produce a perceptually similar output~\cite{horner1993machine,mitchell2007evolutionary}. Sound-matching experiments are either \textit{in-domain} or \textit{out-of-domain} (OOD). For in-domain experiments, the target sound and the mimicking outputs are generated by the same synthesizer. In OOD settings, the target sound originates from a different synthesis process or recording. While OOD sound-matching more closely resembles practical sound design, prior work overwhelmingly focuses on in-domain evaluations, with the goal of replicating the entire sound~\cite{horner1993machine,yee2018automatic,han2023perceptual,vahidi2023mesostructures,salimi2025evaluating}. Since manual audio evaluations are impractical at scale, the lack of reliable automatic evaluation methods is a major hurdle in OOD sound-matching. 

We propose an approach to OOD experiments that can be automatically evaluated, and use this methodology to re-examine the results from recent in-domain experiments. In particular, recent work by Salimi et al.~\cite{salimi2025evaluating} tested audio loss functions (differentiable measures of similarity between two sounds) in different in-domain sound-matching tasks and observed that there is no best universal loss function; that is, the performance of loss functions depends on the method of synthesis. Here, we test whether the reported observations also occur in OOD scenarios, and verify the results with listening tests. Each scenario pairs a target synthesizer with an imitator whose parameters are optimized to mimic the target's output.

The synthesizers are designed to mismatch while retaining a partial, critical overlap in parameters. In addition, the synthesizers are minimal representations of common signal processing functions used by sound designers (e.g., filtering, amplitude/pitch modulation)~\cite{salimi2025evaluating}. This also simplifies the selection of a critical parameter for each synthesizer pair. For example, if two synthesizers apply amplitude modulation to different source signals, the modulation rate is the partially overlapping parameter and a reasonable starting point for measuring success. Matching a target sound's critical characteristics, rather than replicating it outright, mirrors the practice of sound design; success on critical parameters is thus a meaningful unit of progress for OOD sound-matching. We call the application of parameter loss to mismatched synthesizers \textit{partial parameter distance} (PPD). 

\textit{Our central hypothesis is that the performance rankings of loss functions vary with the synthesizers used in OOD scenarios.} We address three research questions: \textbf{(RQ1)} does the synthesis-dependent loss behavior observed in-domain also occur in OOD settings? \textbf{(RQ2)} do the reported advantages and limitations of common loss functions persist under partial synthesizer mismatch? \textbf{(RQ3)} can PPD serve as an automatic evaluation procedure for OOD sound-matching, and how well does it align with human perception? To answer these questions, we contribute a methodology for controlled differentiable OOD sound-matching experiments based on synthesizer pairs with partially overlapping functionality, with PPD as its automated evaluation procedure, validated against blinded listening tests; the framework can serve as a screening tool for certifying loss functions before they are deployed in more complex settings. Replication code will be released upon publication.

\section{Background}
\label{sec:background}

Sound-matching optimizes parameters $\theta$ of a synthesizer $g(\theta)$ so that $x_\theta = g(\theta)$ matches a target $x^t = g^t(\theta^t)$ under a similarity measure $L(x_\theta, x^t)$; in a differentiable pipeline, $\theta$ is updated via gradients of $L$~\cite{yee2018automatic,esling2019flow}. Most similarity measures fall into two families: audio-based and parameter-based. Audio-based losses compute distances between acoustic representations such as STFT spectrograms (e.g., L1, L2, or scale-invariant MSE) or amplitude envelopes; extensions such as Joint Time-Frequency Scattering (JTFS) apply wavelet transformations to a time-frequency representation and have been proposed for capturing \textit{mesostructures}~\cite{anden2015joint,vahidi2023mesostructures,lostanlen2019shape}. Parameter Loss (P-Loss) compares the target and imitator parameter vectors directly (e.g., $\|\theta - \theta^t\|_2$) and requires that both synthesizers share the same parameter space. While P-Loss can serve as a training objective when ground-truth parameters are available~\cite{yee2018automatic,esling2019flow}, it is undefined in OOD settings, where there is no 1-to-1 parameter mapping between the target and imitator synthesizers.

Systematic comparison of loss functions under controlled OOD conditions is largely absent from the literature; most works evaluate exclusively in-domain~\cite{horner1993machine,mitchell2007evolutionary,yee2018automatic,han2023perceptual,vahidi2023mesostructures,barkan2019inversynth}. Recent OOD work has focused on training-time adaptation: fine-tuning parameter estimators on OOD targets~\cite{masuda2021soundmatch,masuda2023improving,uzrad2024diffmoog} or using reinforcement learning for cross-domain matching~\cite{shin2025synthrl}, rather than comparing loss functions in controlled OOD conditions. We address this gap by comparing loss functions across OOD scenarios built from minimal synthesizer pairs, each representing a fundamental approach to sound design~\cite{salimi2025evaluating}.

\section{Methodology}
\label{sec:methodology}
\subsection{Problem Setup}
We evaluate four differentiable loss functions across seven scenarios (six OOD, one in-domain). A \textbf{scenario} is an imitator--target synthesizer pair $(g, g^t)$; for each scenario we run one experiment per loss function, and each experiment consists of at least 300 independent trials in which the target parameters $\theta^t$ and the imitator parameters $\hat{\theta}$ are initialized by uniform sampling over predefined ranges; 200 steps of gradient-based optimization then update $\hat{\theta}$ to minimize the loss between $g(\hat{\theta})$ and $g^t(\theta^t)$. This yields at least 300 PPD scores per loss per scenario for statistical comparison; listening tests provide verification. All synthesizers produce 1-second signals at 48{,}000~Hz.

\subsection{Loss Functions}
\label{subsec:loss_functions}
The four loss functions are as follows. \LoneSpec is the pointwise L1 distance between log-magnitude STFT spectrograms. \SIMSESpec replaces the L1 term with Scale-Invariant Mean Squared Error (SIMSE)~\cite{barron2014shapessimse}, which is invariant to constant offsets in log-magnitude and therefore less sensitive to global level differences. \JTFS applies the L2 distance to the 1-D Joint Time-Frequency Scattering transform~\cite{anden2015joint} ($J{=}10$, $Q{=}2$), which captures wavelet-based time-frequency structure. \DTWEnv computes the soft-DTW distance~\cite{cuturi2017soft} ($\gamma{=}1$) between amplitude envelopes, where the envelope at each time step is obtained by summing all frequency bins of an STFT. For all STFT-based computations (\LoneSpec, \SIMSESpec, and the \DTWEnv envelope), we use a 512-point FFT (257 frequency bins), a Hann window, and a hop length of 100 samples.

\subsection{Evaluation Methods}
\label{subsec:evaluation}
\noindent\textbf{Partial Parameter Distance.}\label{sec:partial_ploss} In OOD settings, full parameter comparison is undefined because the target and imitator synthesizers have different architectures or parameter ranges. For example, if the target is a sine-wave AM synthesizer (carrier 200~Hz, modulation 10~Hz) and the imitator is a saw-wave AM synthesizer (carrier 30--5000~Hz, modulation 1--15~Hz), the only reasonable point of comparison is the modulation rate. \textit{Partial Parameter Distance} (PPD) applies P-Loss only to the subset of parameters with critical perceptual correspondence between the two synthesizers, computed after optimization:
\begin{equation}
\mathrm{PPD} = \| \theta^{t}_{\mathrm{crit}} - \hat{\theta}_{\mathrm{crit}} \|_2,
\end{equation}
where $\theta_{\mathrm{crit}}$ denotes normalized $[0,1]$ parameters common to $g$ and $g^t$. Selecting critical parameters requires domain knowledge and carries some subjectivity; our scenarios (Section~\ref{sec:scenarios}) are designed with a single dominant synthesis task per pair to keep this selection unambiguous. PPD is thus a per-scenario evaluation procedure rather than a universal metric, consistent with the premise that no single similarity measure serves all synthesis tasks~\cite{salimi2025evaluating}.

\noindent\textbf{Listening tests.} Listening tests serve as the primary ground truth. Following Salimi et al.~\cite{salimi2025evaluating}, in each scenario we randomly sample 40 (target, imitator) audio pairs per loss function; four authors independently rate each pair on a 5-point Likert scale (``are the two sounds similar?''), yielding 160 scores per loss per scenario. Raters are blinded to the generating loss function, and three of the four raters were not involved in synthesizer implementation, limiting familiarity-driven bias.

\subsection{Optimization Loop}
\label{subsec:optimization}
The four loss functions are implemented as differentiable JAX functions, and the synthesizers are written in Faust and transpiled into JAX via DawDreamer~\cite{braun2021dawdreamer}; this pipeline is faster than real-time and parallelizable. Parameters are updated using RMSProp with a fixed learning rate of 0.045 for 200 iterations, following~\cite{salimi2025evaluating}. Gradients are clipped to an L2 norm of 1 to mitigate exploding gradients~\cite{salimi2025evaluating}.

\subsection{Bootstrapping and Ranking Procedures}
\label{sec:bootstrapping_ranking}
We follow the methodology of Salimi et al.~\cite{salimi2025evaluating} for ranking the loss functions. For each loss function and scenario, we bootstrap 1,000 resampled means from the PPD scores (or 160 Likert scores) to obtain nonparametric performance distributions~\cite{chernick2011bootstrap}; since PPD is a distance, we use the reciprocal of each resampled mean (1/PPD, as plotted in Fig.~\ref{fig:npsk_merged}) so that higher values indicate better performance for both measures. These distributions are then ranked using the \textit{nonparametric Scott-Knott procedure} (NPSK), which recursively partitions distributions into statistically distinct groups without assuming normality~\cite{tantithamthavorn2017mvt}. Rank 1 indicates the best-performing loss; ties indicate distributions that NPSK cannot distinguish. The scenario rankings are summarized in Figure~\ref{fig:npsk_merged}; we discuss each scenario and its rankings in the following section.

\section{Scenarios and Results}
\label{sec:scenarios}
\begin{figure*}[!t]
  \centering
  \includegraphics[width=\textwidth]{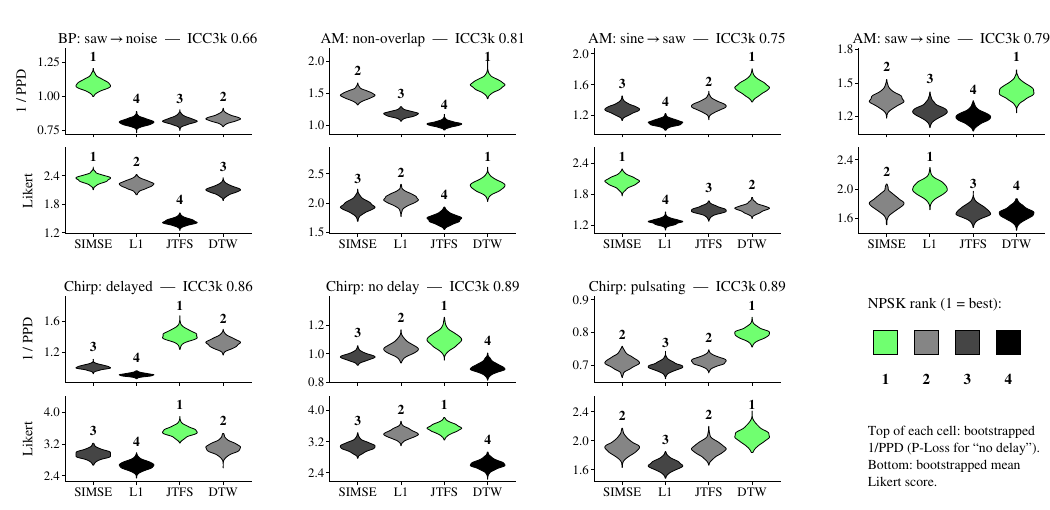}
  \caption{Bootstrapped $1/\mathrm{PPD}$ (top panel of each cell; P-Loss for the in-domain ``no delay'' scenario) and mean Likert rating (bottom panel) distributions for each scenario, with violins colored and numbered by NPSK rank (1 = best). Higher is better on both axes. Scenario labels read target $\rightarrow$ imitator; each title gives the ICC3k inter-rater reliability of the four-rater average.}
  \label{fig:npsk_merged}
\end{figure*}

The scenarios tested here use simple synthesis methods common in
prior work~\cite{engel2020ddsp,vahidi2023mesostructures} and sound design~\cite{roads1996computer}; this simplicity isolates the interaction between loss function and synthesizer~\cite{salimi2025evaluating}. Searchable parameters are likewise restricted to those that dominate each sound's perceptual identity (e.g., filter cutoffs rather than resonance), keeping critical-parameter selection unambiguous; the pulsating chirp is deliberately uncommon, composing a pitch trajectory with an LFO to probe \JTFS's reported strength on mesostructures~\cite{vahidi2023mesostructures}.
Figure~\ref{fig:npsk_merged} shows the per-scenario bootstrapped distributions, NPSK ranks, and inter-rater reliability.

\subsection{Band-Pass Matching}
\label{sec:scenario_bp_saw_noise}
Two synthesizers apply band-pass filtering to different source signals: \BPSaw (target: band-passed saw oscillator) and \BPNoise (imitator: band-passed white noise). Both share cutoffs \texttt{lp\_cut}$\in$[100,5000]~Hz and \texttt{hp\_cut}$\in$[1,400]~Hz; the differing source signals (broadband stochastic vs.\ deterministic harmonic series) make exact replication impossible. PPD is the L2 distance between the normalized cutoffs. Both PPD and listeners rank \SIMSESpec first (Fig.~\ref{fig:npsk_merged}), consistent with in-domain findings that spectrogram-based losses are best for filter-cutoff recovery~\cite{salimi2025evaluating}; the remaining ranks disagree between the two measures.

\subsection{AM-Synthesizer Matching}
\label{sec:am_sound_matching}
Three OOD scenarios are built from two AM synthesizers: \FMMod (saw carrier modulated by a sine LFO) and \FMModvtwo (sine carrier modulated by a sine LFO). Each has a modulation-rate parameter \texttt{amp}$\in$[1,15]~Hz and a carrier frequency \texttt{carrier}. Because the modulator sits in the sub-audio range, the effect is a slow tremolo rather than the sidebands of true ring modulation. PPD is the L2 distance on \texttt{amp}; \texttt{carrier} is excluded because the carrier ranges or waveforms differ by design and, given logarithmic pitch perception, no single ``correct'' frequency match exists~\cite{young1939terminology}. The three variants are: \textit{(i) non-overlapping frequencies}: two \FMModvtwo instances with disjoint carrier ranges (50--1500~Hz target, 2000--8000~Hz imitator); \textit{(ii) sine$\to$saw} and \textit{(iii) saw$\to$sine}: mirrored scenarios between \FMModvtwo and \FMMod on overlapping carrier ranges (30--5000~Hz), where numerical carrier values can coincide but harmonic content still differs. \DTWEnv ranks first on PPD in all three, confirming accurate recovery of the modulation rate. Listeners agree in the non-overlapping case, but in the overlapping-carrier scenarios they prefer a spectrogram-based loss, whose outputs achieve better timbral similarity at the cost of envelope accuracy. This highlights the divergence between PPD and holistic perceptual judgments when multiple perceptual factors compete.
\subsection{Pitch-Bending}
\label{sec:chirplet_scenarios}
Three ``chirplet''~\cite{mann1995chirplet,lostanlen2019shape} scenarios use a base synthesizer that emits a sine tone at a starting pitch (30--100~Hz) that, after $\delta$ samples of steady tone, undergoes an exponential upward pitch bend at rate \texttt{increase\_speed}$\in$[1,20]. (i) \textit{Delayed} (OOD): $\delta$ is drawn uniformly from 5000--30000 samples and is not searched, so the imitator cannot replicate the onset time; PPD is the L2 distance on \texttt{starting\_pitch} and \texttt{increase\_speed}. (ii) \textit{No-delay} (in-domain, $\delta=0$): a close replication of the chirplet experiments in Vahidi et al.~\cite{vahidi2023mesostructures}. (iii) \textit{Pulsating} (OOD): a saw-wave LFO with frequency \texttt{pulse\_rate}$\in$[1,10]~Hz oscillates the pitch around the increasing trajectory (\texttt{increase\_speed} restricted to $[2,7]$ for this variant); the pulse depth is an unsearched constant sampled per synthesizer (50--1000~Hz), so target and imitator generally cannot coincide. PPD is the L2 distance on \texttt{increase\_speed} and \texttt{pulse\_rate}. \JTFS ranks first in the delayed and no-delay scenarios under both parameter-based and listening-test rankings, consistent with its reported agnosticism to small time and frequency shifts~\cite{vahidi2023mesostructures}. In the pulsating scenario, \DTWEnv displaces \JTFS to a tie for second place with \SIMSESpec.

\section{Discussion}
\label{sec:discussion}

We find that synthesis-dependent loss behavior, as reported in previous in-domain work~\cite{salimi2025evaluating}, generalizes to OOD settings \textbf{(RQ1)}: \SIMSESpec is best for band-pass matching, \DTWEnv for amplitude modulation and pulsating pitch, and \JTFS for smooth pitch trajectories. \LoneSpec ranks first in only one case (Likert on AM: saw$\to$sine) and is assigned NPSK rank 4 in five of the fourteen rankings of Fig.~\ref{fig:npsk_merged}, while \SIMSESpec is never ranked last, suggesting that scale invariance makes spectrogram losses more robust across synthesis tasks \textbf{(RQ2)}. PPD agrees with listening tests on the top-ranked loss in five of seven scenarios \textbf{(RQ3)}; the two disagreements occur in AM with overlapping carriers (Section~\ref{sec:am_sound_matching}), where \DTWEnv ranks first on PPD (accurate modulation rate) but listeners favor spectrogram losses (better timbral similarity). This emphasizes that PPD measures a targeted parameter set and should be paired with listening tests when multiple perceptual factors compete. Notably, \JTFS excels on smooth pitch bends but is displaced by \DTWEnv on the pulsating variant \textbf{(RQ2)}, indicating that its ``mesostructure'' domain is more narrowly scoped than previously suggested~\cite{salimi2025evaluating,vahidi2023mesostructures}. Inter-rater reliability of the four-rater average used in our analysis, measured by the intraclass correlation coefficient for the mean of $k{=}4$ raters (ICC3k)~\cite{liljequist2019intraclass}, is good (pooled ICC3k = 0.86, $p < 0.001$); per-scenario ICC3k values appear in the panel titles of Fig.~\ref{fig:npsk_merged}. The listening-derived rankings are also robust to panel composition: the top-ranked loss is unchanged under every leave-one-rater-out ablation (mean rank correlation $\rho = 0.98$). The noisiness of individual similarity judgments in OOD settings is itself an argument for complementing listening panels with automatic measures such as PPD. \\ 
\noindent\textbf{Practical implications.} Loss selection should follow the synthesis task: spectrogram-based losses for filtering (with \SIMSESpec preferred to \LoneSpec), \DTWEnv for amplitude modulation, and \JTFS for smooth pitch trajectories. For OOD evaluation, PPD with overlapping synthesizers is a viable strategy provided the critical parameters are perceptually dominant; pairing with a small listening panel is recommended. PPD is not intended for arbitrary synthesizer pairs, where critical parameters may be ill-defined; rather, it serves as a screening tool: loss functions certified in controlled scenarios can then be deployed in complex settings where parameter correspondence no longer exists. PPD may also complement P-Loss in in-domain settings where the goal is imitation of specific sound characteristics rather than direct parameter replication.

\noindent\textbf{Limitations and future work.} Whether PPD scales to higher-dimensional parameter spaces or generalizes to real recorded targets is an open question; the selection of critical parameters also carries subjectivity that our simple synthesizer pairs sidestep. Our listening panel is small and composed of authors; blinding and high inter-rater reliability mitigate but do not remove this limitation, and validation against a larger external panel remains future work. All losses also share a single optimizer configuration, so rankings may partly reflect interactions between loss scaling and the learning rate rather than intrinsic loss behavior. Our results are also conditioned on gradient-based optimization: differentiable pipelines are known to struggle with non-convex loss surfaces induced by oscillatory parameters~\cite{turian2020sorry,hayes2023sinusoidal,salimi2025evaluating}, so loss rankings may shift under other search strategies (e.g., evolutionary methods). Promising directions include OOD matching with pre-recorded targets, real-time sound-matching~\cite{shier2024real}, and composite matching algorithms that deploy task-specific losses validated in isolation (e.g., \DTWEnv for envelope recovery and \JTFS for pitch trajectories) within a single system.
\bibliographystyle{IEEEtran}
\bibliography{references}

@article{vahidi2023mesostructures,
  author={Vahidi, Cyrus and Han, Han and Wang, Changhong and Lagrange, Mathieu and Fazekas, Gy{\"o}rgy and Lostanlen, Vincent},
journal={J. Audio Eng. Soc.},
title={Mesostructures: Beyond spectrogram loss in differentiable time--frequency analysis},
year={2023},
volume={71},
number={9},
pages={577-585},
month={September},}

@inproceedings{anden2015joint,
  title={Joint time-frequency scattering for audio classification},
  author={And{\'e}n, Joakim and Lostanlen, Vincent and Mallat, St{\'e}phane},
  booktitle={Proc. IEEE MLSP},
  pages={1--6},
  year={2015},
}

@inproceedings{engel2020ddsp,
  title={{DDSP}: Differentiable digital signal processing},
  author={Engel, Jesse and Hantrakul, Lamtharn and Gu, Chenjie and Roberts, Adam},
  booktitle={Proc. ICLR},
  year={2020}
}

@inproceedings{masuda2021soundmatch,
  title={Synthesizer Sound Matching with Differentiable {DSP}},
  author={Masuda, Naotake and Saito, Daisuke},
  booktitle={ISMIR},
  pages={428--434},
  year={2021}
}

@article{masuda2023improving,
  title={Improving semi-supervised differentiable synthesizer sound matching for practical applications},
  author={Masuda, Naotake and Saito, Daisuke},
  journal={IEEE/ACM Trans. Audio, Speech, Lang. Process.},
  volume={31},
  pages={863--875},
  year={2023},
  publisher={IEEE},
  doi={10.1109/TASLP.2023.3237161}
}

@article{uzrad2024diffmoog,
  title={{DiffMoog}: A Differentiable Modular Synthesizer for Sound Matching},
  author={Uzrad, Noy and Barkan, Oren and Elharar, Almog and Shvartzman, Shlomi and Laufer, Moshe and Wolf, Lior and Koenigstein, Noam},
  journal={arXiv preprint arXiv:2401.12570},
  year={2024}
}

@inproceedings{han2023perceptual,
  title={Perceptual--neural--physical sound matching},
  author={Han, Han and Lostanlen, Vincent and Lagrange, Mathieu},
  booktitle={Proc. IEEE ICASSP},
  pages={1--5},
  year={2023},
  organization={IEEE}
}

@article{esling2019flow,
  title={Flow synthesizer: Universal audio synthesizer control with normalizing flows},
  author={Esling, Philippe and Masuda, Naotake and Bardet, Adrien and Despres, Romeo and Chemla-Romeu-Santos, Axel},
  journal={Applied Sciences},
  volume={10},
  number={1},
  pages={302},
  year={2019},
  publisher={MDPI}
}

@inproceedings{mitchell2007evolutionary,
  title={Evolutionary sound matching: A test methodology and comparative study},
  author={Mitchell, Thomas J and Creasey, David P},
  booktitle={Proc. ICMLA},
  pages={229--234},
  year={2007},
}

@article{horner1993machine,
  title={Machine tongues {XVI}: Genetic algorithms and their application to {FM} matching synthesis},
  author={Horner, Andrew and Beauchamp, James and Haken, Lippold},
  journal={Computer Music Journal},
  volume={17},
  number={4},
  pages={17--29},
  year={1993},
  publisher={JSTOR}
}

@article{yee2018automatic,
  title={Automatic programming of {VST} sound synthesizers using deep networks and other techniques},
  author={Yee-King, Matthew John and Fedden, Leon and d'Inverno, Mark},
  journal={IEEE Trans. Emerg. Topics Comput. Intell.},
  volume={2},
  number={2},
  pages={150--159},
  year={2018},
  publisher={IEEE},
  doi={10.1109/TETCI.2017.2783885}
}

@article{turian2020sorry,
  title={I'm sorry for your loss: Spectrally-based audio distances are bad at pitch},
  author={Turian, Joseph and Henry, Max},
  journal={arXiv preprint arXiv:2012.04572},
  year={2020}
}

@inproceedings{cuturi2017soft,
  title={Soft-{DTW}: A differentiable loss function for time-series},
  author={Cuturi, Marco and Blondel, Mathieu},
  booktitle={Proc. ICML},
  pages={894--903},
  year={2017},
  organization={PMLR}
}

@article{barron2014shapessimse,
  title={Shape, illumination, and reflectance from shading},
  author={Barron, Jonathan T and Malik, Jitendra},
  journal = {IEEE Trans. Pattern Anal. Mach. Intell.},
  volume={37},
  number={8},
  pages={1670--1687},
  year={2015},
  publisher={IEEE}
}

@book{chernick2011bootstrap,
  title={Bootstrap methods: A guide for practitioners and researchers},
  author={Chernick, Michael R},
  year={2011},
  publisher={John Wiley \& Sons}
}

@book{roads1996computer,
  title={The Computer Music Tutorial},
  author={Roads, Curtis},
  year={1996},
  publisher={{MIT Press}}
}

@inproceedings{krekovic2019insights,
  title={INSIGHTS IN HABITS AND ATTITUDES REGARDING PROGRAMMING SOUND SYNTHESIZERS: A QUANTITATIVE STUDY},
  author={Krekovic, Gordan},
  booktitle={Proc. Sound and Music Computing Conf.},
  year={2019}
}

@article{tantithamthavorn2017mvt,
    Author={Tantithamthavorn, Chakkrit and McIntosh, Shane and Hassan, Ahmed E. and Matsumoto, Kenichi},
    Title = {An Empirical Comparison of Model Validation Techniques for Defect Prediction Models},
    Journal = {IEEE Trans. Softw. Eng.},
    Volume = {43},
    Number = {1},
    pages = {1--18},
    Year = {2017}
}

@inproceedings{braun2021dawdreamer,
  title={{DawDreamer}: Bridging the Gap Between Digital Audio Workstations and {Python} Interfaces},
  author={Braun, David},
  booktitle={Late-Breaking Demo, ISMIR},
  year={2021}
}

@article{barkan2019inversynth,
  title={Inversynth: Deep estimation of synthesizer parameter configurations from audio signals},
  author={Barkan, Oren and Tsiris, David and Katz, Ori and Koenigstein, Noam},
  journal={IEEE/ACM Trans. Audio, Speech, Lang. Process.},
  volume={27},
  number={12},
  pages={2385--2396},
  year={2019},
  publisher={IEEE}
}

@article{salimi2025evaluating,
  title={Evaluating Sound Similarity Metrics for Differentiable, Iterative Sound-Matching},
  author={Salimi, Amir and Hindle, Abram and Za{\"\i}ane, Osmar R},
  journal={IEEE Trans. Audio, Speech, Lang. Process.},
  volume={33},
  pages={4982--4994},
  year={2025},
  publisher={IEEE}
}

@article{young1939terminology,
  title={Terminology for logarithmic frequency units},
  author={Young, Robert W},
  journal={J. Acoust. Soc. Am.},
  volume={11},
  number={1},
  pages={134--139},
  year={1939},
  publisher={Acoustical Society of America}
}

@inproceedings{lostanlen2019shape,
  title={The shape of RemiXXXes to come: audio texture synthesis with time-frequency scattering},
  author={Lostanlen, Vincent and Hecker, Florian},
  booktitle={Proc. DAFx},
  year={2019}
}

@inproceedings{shin2025synthrl,
  title={{SynthRL}: Cross-domain Synthesizer Sound Matching via Reinforcement Learning},
  author={Shin, Wonchul and Lee, Kyogu},
  booktitle={Proc. IJCAI},
  pages={10162--10170},
  year={2025}
}

@article{mann1995chirplet,
  title={The chirplet transform: Physical considerations},
  author={Mann, Steve and Haykin, Simon},
  journal={IEEE Trans. Signal Process.},
  volume={43},
  number={11},
  pages={2745--2761},
  year={1995},
  publisher={IEEE}
}

@article{liljequist2019intraclass,
  title={Intraclass correlation--A discussion and demonstration of basic features},
  author={Liljequist, David and Elfving, Britt and Skavberg Roaldsen, Kirsti},
  journal={PloS one},
  volume={14},
  number={7},
  pages={e0219854},
  year={2019},
  publisher={Public Library of Science San Francisco, CA USA}
}

@inproceedings{hayes2023sinusoidal,
  title={Sinusoidal frequency estimation by gradient descent},
  author={Hayes, Ben and Saitis, Charalampos and Fazekas, Gy{\"o}rgy},
  booktitle={Proc. IEEE ICASSP},
  pages={1--5},
  year={2023},
}

@article{shier2024real,
  title={Real-time timbre remapping with differentiable {DSP}},
  author={Shier, Jordie and Saitis, Charalampos and Robertson, Andrew and McPherson, Andrew},
  journal={arXiv preprint arXiv:2407.04547},
  year={2024}
}


\end{document}